\documentclass{elsart}

\usepackage{epsfig}
\usepackage{amssymb}
\begin{document}
\begin{frontmatter}
\title{Evaluation of the ultimate performances
of a Ca$^+$ single-ion frequency standard}
\author{C. Champenois}, \ead{caroline.champenois@up.univ-mrs.fr} \author{M. Houssin}, \author{C. Lisowski
}, \author{ M. Knoop}, \author{G. Hagel }, \author{ M. Vedel }  and 
\author{ F. Vedel} \address{Physique des interactions
ioniques et mol\'eculaires
\\ Unit\'e mixte de recherche 6633 CNRS-Universit\'e de Provence\\
Centre de Saint-J\'er\^ome, Case C21, 13397 Marseille Cedex 20, France.}

\date{\today}

\begin{abstract}
We numerically evaluate the expected performances of an optical
frequency standard at 729 nm based on a single calcium ion. The
frequency stability is studied through the Allan deviation and its
dependence on the excitation method (single Rabi pulse or two Ramsey pulses schemes) and the laser linewidth are discussed.
 The minimum Allan deviation that can be expected is estimated to
 $\sigma_y(\tau) \approx 2.5\times 10^{-15}/\sqrt{\tau}$ with $\tau$ the integration time. The
  frequency shifts induced by the environmental conditions are evaluated
  to minimize the uncertainty of the proposed standard by chosing
  the most suited environment  for the ion.
 If using the odd isotope $^{43}$Ca$^{+}$ and a vessel cooled to 77 K, the expected relative shift  is  $-2 \times 10^{-16}$ with an uncertainty of
$\pm 4\times10^{-16}$, mainly due to the quadrupole shift induced by the unknown static electric field gradient .
\end{abstract}

\end{frontmatter}
\section{Introduction}
\label{s-introduction}  Thanks to the recent progress
made in atom and ion cooling and trapping, laser stabilization
and high-resolution optical spectroscopy, narrow optical transitions are considered
as a  basis for frequency standards. At this time, two kinds of
experiments are under study in various groups: one uses  an
ensemble of laser-cooled neutral atoms in a fountain, an optical
lattice or a BEC, the other one uses a single trapped laser-cooled
ion (for a recent review see \cite{braun03}). The present work is
motivated by strong progress in storing, cooling and
coherently manipulating single ions in Paul traps. Together with
the ultra-precise optical frequency measurements achieved by
frequency chains and frequency combs, these progress lead to
 the realization of single-ion frequency standards, as  for
Hg$^+$\cite{rafac00}, Sr$^+$\cite{marmet00,gill03},
Yb$^+$\cite{gill03} and In$^+$\cite{eichenseer03}, and proposed
for Ca$^+$ \cite{cc01}. Our experimental project aims to
build an optical frequency standard using the electric quadrupole
transition $4 S_{1/2}\rightarrow 3 D_{5/2}$ of a single calcium
ion at 729 nm. Among the frequency standard candidates, Ca$^{+}$
possesses the major advantage that the required radiations for
cooling and exciting the clock transition can be produced directly by
solid state or diode lasers. In addition, the existence of an isotope having semi-integer nuclear spin ($I=7/2$) allows to eliminate the first order Zeeman shift, canceling a major source of line shift and broadening.

The performances of a frequency standard are defined by the
stability of its local oscillator (a laser in the optical case)
and  the precision achieved in the observation of an atomic
transition.  Frequency instability is due to deviations from a
mean frequency throughout varying probe time intervals, while
frequency uncertainty is caused by the atomic frequency fluctuations
induced by environmental conditions and
 by the experimental conditions for observation. The quality factor  of
 optical atomic transitions  can
reach 10$^{15}$, which is 5 orders of magnitude higher than for
microwave frequency standards and thus let hope  better ultimate performances than
the existing standards.

The interrogation scheme used to probe the atomic transition influences the frequency stability of the proposed standard via the variation of the duration of the probe cycle.
In the first part of this  paper we discuss the choice of this scheme, by using  numerical simulations to
compare single-pulse spectroscopy with time-domain Ramsey
interferometry. Systematic effects expected
for a standard based on $^{43}$Ca$^+$  may reduce its
accuracy and precision, they  are discussed in the second part. For this evaluation, we
 employ the specific parameters of the Ca$^+$-ion trap
experiment in Marseille \cite{cc01} as an example, but the
discussion is kept as general as possible to remain applicable to other
atomic species.

\section{Frequency stability\label{s-stabilite}}
Frequency stability is one of the major characteristics of a
frequency standard. It can be quantified by the Allan deviation
$\sigma_y(\tau)$ measured for an average time $\tau$:
\begin{equation}
 \sigma_y(\tau)=\frac{1}{Q S/N}\sqrt{\frac{T_c}{\tau}}
\end{equation}
where $Q=f_0/\Delta f$ is the quality factor defined by the ratio
of the clock frequency over its observed linewidth, $S/N$ the
signal to noise ratio and $T_c$ the cycle time required for the
interrogation of the ion.

\begin{figure}
\begin{center}
 \includegraphics[height=5.cm]{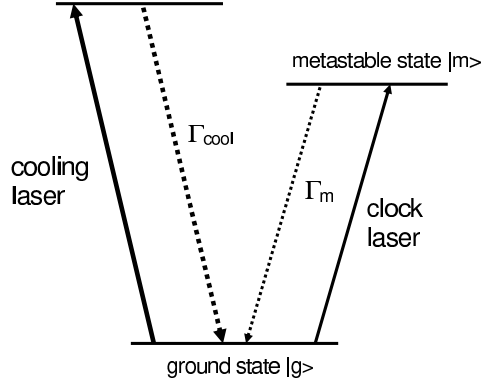}
 \caption{Transitions involved in the cooling and probing of an
 ion for a frequency standard. In practice, the cooling scheme may
 involve several lasers because of a possible hyperfine structure
 and/or a possible decay towards other metastable states. In most
 cases, $\Gamma_{cool}$ is of the order of 20 MHz whereas
 $\Gamma_m$ is of the order of 1 Hz.}
  \label{fig_lasers}
\end{center}
 \end{figure}

The schemes to probe  an optical transition of a single ion
consist  of a preparation stage, an excitation of the clock
transition, and a final detection stage. The cycle time $T_c$ is
the sum of the corresponding time durations $T_{prep}$,
$T_{exc}$, and $T_{det}$. During the preparation stage, the ion is laser
cooled and optically pumped into the internal state
chosen to be the ground state (see figure \ref{fig_lasers}).
 The light scattered by the
  cooling transition is also used for detection of the ion. The probing of
 the clock transition by the local oscillator (a
laser) is done by direct laser excitation. During this stage, all
the preparation lasers are shut off. After excitation of the clock
transition, the cooling lasers are switched on again.  The
quantum jump method  \cite{dehmelt67} allows then to
know if the atom is excited or not: the absence of fluorescence
  during the detection stage proves that the ion is
"shelved" in the metastable upper state whereas the presence of
fluorescence signal means that the ion is still in the cooling
cycle. Repetition of this measurement as a function of the clock
laser frequency allows to measure the transition probability
distribution.  In the following, we discuss the laser characteristics and the maximum duration required for these various
stages. The choice of the suited interrogation scheme is essential to minimize
the Allan deviation.

\subsection{Preparation and detection\label{s-prep}}
 We suppose the ion cooled to the Doppler limit in an RF-trap and located at the center of the trap,
where the RF-trapping field is minimum and   has little influence
on the ion motion. Since the trap creates a quasi-harmonic potential,
the motion of the ion is a superposition of oscillations at
different frequencies due to the spatial anisotropy of the
trapping device. If we suppose, for simplicity's sake, that there
is only one frequency of motion $\omega_{tr}/2\pi$ in the trap,
the resulting atomic absorption spectrum is composed of a central
frequency corresponding to the atomic transition $\omega_0/2\pi$
and sidebands separated by multiples of the motional frequency
$(\omega_0 \pm p\omega_{tr})/2\pi$, ($p$ integer). The sidebands
are resolved if the width of each band is smaller than their
mutual separation (the strong confinement condition
\cite{wineland79}). This can be  achieved in miniature
traps with high motional frequencies ($\approx 1$ MHz) for all
the narrow transitions considered as potential basis for frequency
standards ($\Gamma_m\ll\omega_{tr}$, see figure
\ref{fig_lasers}). The intensity of each band $p$ in the spectrum
depends on the oscillation amplitude $X$ of the ion in the trap
like $J_p^2(kX)$ \cite{wineland79} where $k$ is the laser
wavevector of the probed transition and $J_p$ the Bessel function
of order $p$. As the functions $J_p^2(kX)$ have negligeable values
when $p \gtrsim kX$, the smaller is $kX$, the less sidebands
are visible. Laser-cooling the ion reduces its oscillation
amplitude $X$ and thus the number of observable sidebands. A
major step in the preparation of the ion is to access the
Lamb-Dicke regime which is characterized by the reduction of the
spectrum to few sidebands with a preponderant  weight on
the central frequency, this regime is reached if $kX\lesssim 1$.

The motion of the ion can be described by the occupation rate of the vibrational quantum levels,
characterised by the mean vibrational quantum number $\left<n\right>$.   This
 vibrational state can also be characterized from the classical point of
view by an oscillation amplitude $X=\lambda_{tr}\sqrt{2\langle n\rangle +1}
$, where the length $\lambda_{tr} = \sqrt{\hbar/2m
\omega_{tr}}$ measures the size of the fundamental harmonic
oscillator eigenstate $\left| n=0\right>$. This length depends on the atomic mass by  $1/\sqrt{m}$ and, as an example, is equal to 11 nm for a
calcium ion with $\omega_{tr}=2\pi\times 1$ MHz. The Lamb-Dicke
condition $kX\lesssim 1$ can also be expressed by $k\lambda_{tr}
\sqrt{2\langle n\rangle +1} \lesssim 1$. The Lamb-Dicke parameter
$\eta=k\lambda_{tr}$ quantifies the ability for a system
\{ion+trap\} to reach the Lamb-Dicke regime for a given transition.
 It is of the
order of 0.1 for optical transitions ({\it e.g. } 0.095 for
calcium's clock transition in the trap taken as example). 

In most
cases, the frequency of motion in the trap is of the order of 1
MHz, whereas the atomic dipole transition used for laser cooling
has a width $\Gamma_{cool}$ close to $2\pi \times 20$ MHz. On such
broad transition ($ \Gamma_{cool}/2 \gg \omega_{tr}$) the Doppler
limit for laser cooling can be approximated by the one of a free
atom \cite{wineland79}. This leads to a thermal population of the
harmonic trap vibrational levels characterized by $\langle n\rangle \simeq
\Gamma_{cool}/2\omega_{tr} \simeq 10$. The Lamb-Dicke condition
$\eta \sqrt{2\langle n\rangle +1} \lesssim 1$ is then fulfilled
by the vibrational state   reached by Doppler cooling ($\langle
n\rangle  \simeq 10$). This fulfillment  sets the transition
 free of first order Doppler effect, while the second order
Doppler effect is very small (see section \ref{s-doppler2}).
Furthermore the residual  distribution of occupied vibrational levels still allows to
drive coherent dynamics on the clock transition, as required for
the interrogation schemes and  confirmed numerically in the
following. Since the time needed to reach the Doppler cooling limit is
of the order of milliseconds while the optical pumping is faster
than the millisecond, we can  estimate $T_{prep}$ to 5 ms.

The duration required for the detection stage depends on the
fluorescence signal collected on the strong dipole transition.
For such transitions with a width of $\approx$ 20 MHz, one can
expect at least 10$^4$ counts per second (cps) over a stray light
level of less than 100 cps. In these conditions, 10 ms-periods are
sufficient to acquire enough signal to decide if the atom has
been excited  into the metastable state.  As a consequence,
15 ms is a realistic estimation for the sum of the preparation and detection
contributions to the cycle duration. To this minimum cycle
duration must be added the excitation duration time $T_{exc}$. In
the following subsection, we theoretically study the minimization of this
probe time for different excitation schemes, assuming that the total cycle time $T_c=T_{exc}+15 $ ms.

\subsection{Choice of the excitation scheme}
\subsubsection{Evaluation of the minimum Allan deviation} 
The width $\Delta f$ of the observed transition and its signal to
noise ratio depends on the laser excitation scheme. The choice of a high-frequency clock transition (in the optical
domain) allows to reach smaller Allan deviations than the ones
obtained on  frequency standards in the microwave domain.
Until now, the narrowest optical transition linewidth has been observed
on a Hg$^{+}$ ion \cite{diddams01} and has allowed to reach a relative
frequency stability of $7\times 10^{-15}$ over  1 s averaging.
In this article, we discuss possible excitation schemes independently of the
ion implied. We introduce a reduced Allan deviation $\sigma'
=\sigma_y(\tau) \times f_0 \sqrt{\tau}$ to quantify the expected
frequency stability of the standard, keeping in mind that reduced
Allan deviations between 1 and 10 have already been measured by several groups on
 lasers locked on atomic optical transitions.

Let $p_m$ be the probability  for the ion to be in the metastable
state once excited by the clock laser. The frequency of the transition, to which the clock laser will be locked, is deduced from  the probability $p_m$ measured several times on the low and high
frequency sides of the transition. This method of frequency discrimination requires the excitation probability $p_m$ to be around 0.5, where the slope of the probability distribution is the steepest and the frequency sensitivity is the highest.

 Several sources of noise can limit the signal to noise
ratio. Among these is the quantum projection noise \cite{itano93}
which is dominant once the technical noise has been reduced. The
laser excitation creates a
  linear superposition of the ground ($\left|g\right>$) and metastable ($\left|m\right>$)
   states: $\sqrt{1-p_m}\left|g\right>
   +\sqrt{p_m}\left|m\right>$. During the detection stage,
   the atomic state is projected on one of these
  two atomic states. The variance of such a
  measurement is $p_m(1-p_m)$ and causes  a minimum noise $\sqrt{p_m(1-p_m)}$ on the transition probability. This  can
   be overcome  by using squeezed states \cite{wineland94}, which we do not
   consider here.

   The maximum signal to noise ratio which can then be observed is
\begin{equation}\label{SB}
  S/N = \sqrt{\frac{p_m}{1-p_m}}
\end{equation}
which  is maximum at resonance ($p_m = 1$), where the frequency descrimination is unefficient.  Thus, maximum frequency
sensitivity and maximum signal to noise ratio are not compatible.
Additionally, the finite upper-state lifetime  leads to
spontaneous decay which, for long excitation time,  can reduce the maximum excitation
probability. As a consequence,  finding a compromise between all these incompatible requirements deserve precise studies of  the excitation scheme.

Two excitation schemes have been experimentally tested by several
groups: a single Rabi pulse or two temporally separated Ramsey
pulses. The first one has been performed on Hg$^{+}$
\cite{rafac00}, In$^{+}$ \cite{becker01}, Sr$^{+}$  and
Yb$^{+}$\cite{gill03}, and the second one on Hg$^{+}$
\cite{rafac00} and  Sr$^{+}$ \cite{marmet00}. In the following we
discuss the principal features of each method and then compare
them.
 To quantify the
relative stability allowed by the discussed methods, we evaluate
the reduced Allan deviation by
\begin{equation}\label{sigmaf}
  \sigma^{'}=\sigma_y(\tau)\times f_0 \sqrt{\tau}=\Delta f\sqrt{\frac{1-p_m}{p_m}}\sqrt{T_c}
\end{equation}

For a high frequency sensitivity, we assume that the laser probes the transition on each side of the line,  on the two frequencies corresponding to an excitation probability $p_m$ which is half of the maximum probability measured for zero detuning ($p_m$ can never exceed 1/2).  The deduced $\Delta f$ when these probabilities are equal is then the FWHM of the experimental linewidth.  The evolution of
the density matrix of the two levels  $\left| g \right>$ and
$\left| m \right>$ is computed numerically. The atomic system is defined by the
metastable lifetime $\tau_m$ fixed to 1 second ($\Gamma_m=1/\tau_m$), and it is driven by
a Rabi pulsation $\Omega$. The motion of the ion is taken into account by a distribution of thermal vibrational levels, characterised by the mean vibration number  $\left<n\right>$ and defined by $P(n)=\left( \left<n\right>/(1+\left<n\right>\right)^n$. The excitation probability $p_m$ is then an incoherent weighted sum of the probability for each vibrational level to be excited in the metastable state. For a given laser intensity, the Rabi pulsation from vibrational level $ \left|n\right> \rightarrow  \left|n\right>$ is proportional to $(1-\eta^2n)$ whereas it is proportional to $\eta \sqrt{n+1}$ for a $ \left|n\right> \rightarrow  \left|n+1\right>$ transition. We choose for $\eta$ the value of 0.095 calculated for a calcium ion in the trap described above. As the Doppler cooling leads to $\left<n\right> \simeq 10$, the excitation probability on the $ \left|n\right> \rightarrow  \left|n\right>$ band is higher than the $\left|n\right> \rightarrow \left|n\pm 1\right>$ bands so these last ones were neglected. If $\Omega$ is the Rabi pulsation
for the $\left|n=0\right>\rightarrow \left|n=0\right>$
transition, $\Omega\times L_n(\eta^2)$ is the one for the
$\left|n\right>\rightarrow \left|n\right>$ transition, $L_n$
being the Laguerre polynomial \cite{blockley92}.  The linewidth $\Gamma_L$ of the laser
spectrum (FWHM) is taken into account by adding a source of decoherence
equal to this width in the operator controlling the evolution of
the density matrix \cite{cct_houches75}. For a
laser detuning $\Delta_L$ and in the rotating wave approximation
the density  matrix  evolves according to
\begin{eqnarray}\label{matrice}
\dot{\rho}_{gg} & = &
-\dot{\rho}_{mm}=i\Omega/2\left(\rho_{gm}-\rho_{mg}\right)+\Gamma_m\rho_{mm}
\\
\dot{\rho}_{gm} & = &
-i\Delta_L\rho_{gm}+i\Omega/2\left(\rho_{gg}-\rho_{mm}\right)-\Gamma_m\rho_{gm}/2-\Gamma_L\rho_{gm}/2
\end{eqnarray}

\subsubsection{Single pulse excitation}
To avoid power broadening, the narrow transition can be
experimentally observed if the Rabi pulsation $\Omega$ is smaller
than the linewidth and the interrogation time longer than the
lifetime. But because of the finite lifetime of the excited
state, the maximum excitation probability is low  and requires
several seconds to be reached, reducing the relative stability
even if the observed linewidth is close to the natural width. In an ideal
context where the experiment is not limited by the laser
stability, the calculations show that the smallest reduced Allan
variance is reached with a single pulse which should last at
least 1 s and drive the transition with $\Omega$ of the order of
$2\times \Gamma_m$. With today's laser stability a cycle time
of a few secondes for a single measure seems not realistic. We
rather consider excitation schemes with durations inferior to 1
second, since a cycle has to be repeated several times before a
signal can be built up to counteract on the frequency of the local
oscillator.
 \begin{figure}
\begin{center}
 \includegraphics[height=9.cm]{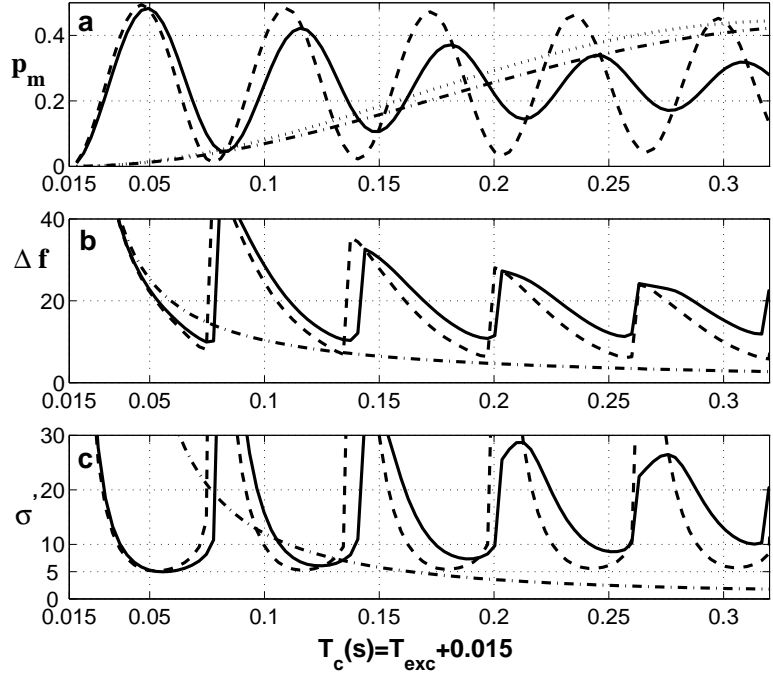}
 \caption{ Calculations of the reduced relative stability $\sigma^{'}$  {\bf (c)}, from the probability at half maximum $p_m$ {\bf (a)} and the width $\Delta f$ (Hz) {\bf (b)} of the excitation probability profile, for an increasing excitation time $T_{exc}$ ($T_c=T_{exc}+0.015$ s). This stability is calculated for a metastable lifetime $\tau_m=1/\Gamma_m=1$ s and a single Rabi pulse ( $\Omega=10 
  \Gamma_m$ or  $\Omega=100  \Gamma_m$ ) for two thermal vibrational states characterised by 
 $\left<n\right>=0$ and  $\left<n\right>=10$.
   Dotted line: $\Omega=10
  \Gamma_m$ and $\left<n\right>=0$, dash-dotted line: $\Omega=10
  \Gamma_m$ and $\left<n\right>=10$, broken line: $\Omega=100  \Gamma_m$  and $\left<n\right>=0
  $, solid line: $\Omega=100 \Gamma_m$  and
  $\left<n\right>=10$.  On {\bf
(b)} and {\bf (c)}, the dotted and dash-dotted line are almost
superimposed and only one curve is plotted on the graph.} \label{fig_rabi}
\end{center}
 \end{figure}

In figure \ref{fig_rabi}, are plotted the excitation probability
at half maximum $p_m$ , the full width at half maximum $\Delta f$ 
 and the reduced stability as defined by equation (\ref{sigmaf}),
versus the cycle time $T_c=T_{exc}+$15 ms, assuming a single Rabi
pulse. These curves reflect Rabi oscillations, which show maximum
excitation probability for $T_{exc}=(2q+1)\pi/\Omega$ ($q$ integer).  For $T_{exc}=2q\pi/\Omega$, the excitation probability is minimum on resonance and shows some maximum for other detunings. The computed FWHM  has then no physical significance, which is not relevant here as a clock is never operated with this excitation duration. The
interesting feature is the  minimum of the reduced Allan variance
observable for the shortest $T_{exc}$. Two cases with different
Rabi pulsation $\Omega$ ($\Omega=10 \Gamma_m$ and
$\Omega=100 \Gamma_m$) are compared in figure
\ref{fig_rabi}. For both cases, excitation probability and full
width at half maximum are computed for an ion whose oscillatory
motion  corresponds to the Doppler
cooling limit ($\left<n\right>=10$) or  to the
fundamental vibrational state ($\left<n\right>=0$). The first
result to mention is that the first minima of the reduced Allan
variance are identical for these two vibrational states, for
the chosen Rabi pulsation. It confirms that Doppler cooling is
sufficient for state preparation. The results shown in figure
\ref{fig_rabi} suggest that the discrepancy between the
$\left<n\right>=10$ and the $\left<n\right>=0$ vibrational state increases with the pulse duration. We have checked
that in the case of the short pulses we consider in the
following, the results are nearly the same for these two
 vibrational distributions and thus, from now on, only the cases
concerning $\left<n\right>=0$ are dealt with.

To illustrate the influence of the
strength of the Rabi pulsation  in figure \ref{fig_rabi}, the excitation is driven by
$\Omega=10 \times \Gamma_m$ and $\Omega=100 \times \Gamma_m$. In
the first case, the minimum reduced Allan deviation is close to 1
but requires an excitation of more than 300 ms. In the second case,
the first minimum is reached for a cycle duration of 50 ms, but
this shortening of the cycle duration is paid by an increase of
$\sigma^{'}$ equal to 4.5. This trend is general and a further
increase of the Rabi pulsation leads to an increase of the
minimum Allan deviation as well as a decrease of the required
cycle time.

\begin{figure}
\begin{center}
 \includegraphics[height=7.cm]{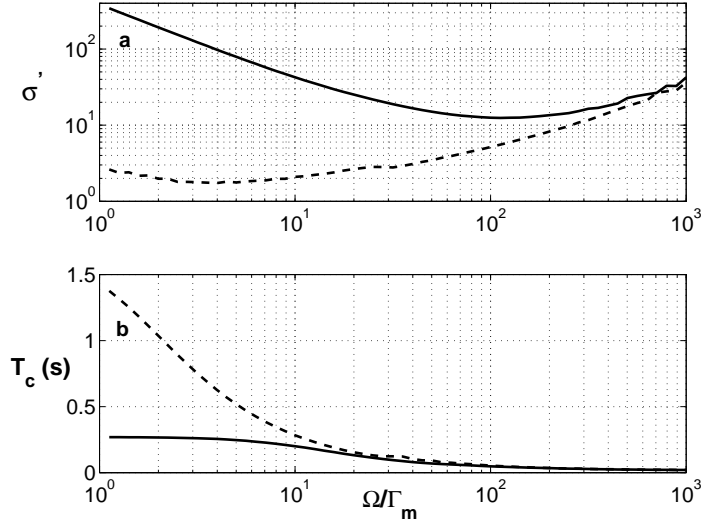}
 \caption{ {\bf (a)}: Minimum reduced Allan deviation and {\bf (b)}:
  cycle duration $T_c$ required to reach this minimum versus the Rabi pulsation.
   Computation done for a metastable lifetime
  $\tau_m=1/\Gamma_m=1$ s and a single Rabi pulse by a laser of width $\Gamma_L$ (FWHM), with an ion in the vibrational ground state ($\left<n\right>=0 $). Broken line: $\Gamma_L/2\pi=0.2$ Hz and
   solid line: $\Gamma_L/2\pi=20$ Hz.
  \label{fig_rabi_broad}}
\end{center}
 \end{figure}

We now take into account the effect of a finite laser linewidth
on  the minimum reduced Allan deviation and compare the value
computed for a laser as broad as the atomic transition
($\Gamma_L/2\pi=0.2$ Hz) to the one computed with a 20
Hz broad laser (FWHM). In this latter case, the performances of the clock
are greatly reduced first by the reduction of the excitation
probability and second by the broadening of the observed
transition.  The first drawback can be overcome by the increase
of the Rabi pulsation but this is paid by a further increase of
the transition broadening. As a consequence, for a given laser
linewidth and a given metastable lifetime, there is an optimal
Rabi pulsation which results in a minimum reduced Allan
deviation. This is illustrated in figure \ref{fig_rabi_broad}
where for $\Gamma_L/2\pi=20$ Hz, $\sigma^{'}$ is minimum ( $\sigma^{'}$=12.4) for
$100\times \Gamma_m \lesssim \Omega \lesssim 125\times \Gamma_m$
 and  a cycle time of 40 ms, whereas
for $\Gamma_L/2\pi=0.2$ Hz, $\sigma^{'}$ is minimum for $\Omega
\simeq 4\Gamma_m$ and is then equal to 1.7 but for a cycle time of
625 ms. In the case of the broadest laser linewidth, these results confirm the intuitive idea that for optimum stability, the pulse length $T_{exc}$ is limited by the laser linewidth $T_{exc}\Gamma_L \simeq \pi$ and that the Rabi pulsation is then approximately set by the resonant $\pi$-pulse condition $\Omega T_{exc}\simeq \pi$. When the laser linewidth is comparable to the transition natural width ($\Gamma_L/2\pi=0.2 $ s$^{-1}$), the optimum stability is reached for a shorter excitation time  $T_{exc}\Gamma_L (\Gamma_m)\simeq 0.6$ and a Rabi pulsation under the resonant $\pi$-pulse condition $\Omega T_{exc}\simeq 2.5$, which could not be dedeuced from the intuitive concept. 
\subsubsection{Comparison with Ramsey interferometry\label{s-ramsey}}
The introduction of the separated fields method or Ramsey
interferometry \cite{ramsey} was soon followed by breakthroughs
in high resolution spectroscopy and is expected to overcome the
limitations met with single pulse excitation. With this method the
line profile is recorded after two pulses of duration $T$ such as
$\Omega T=\pi/2$, separated by a free evolution time $T_{free}$.
When the laser detuning is scanned, the  profile shows Ramsey
fringes resulting from an interference pattern and for short
enough pulse duration (or a high enough Rabi pulsation), the width of the central fringe is equal to
$1/2T_{free}$ and is then independent from the Rabi pulsation.
For a chosen pulsation $\Omega$, the evolution of $p_m$, $\Delta
f$ and $\sigma^{'}$ does not show oscillations with $T_c$ , like
for a single Rabi pulse. $\sigma^{'}$ takes very high values for
short $T_c$ and decreases toward a limit for longer $T_c$. This
limit depends on the choice of the pulsation $\Omega$. 

When the width of the laser is taken into account by the relaxation it causes on the coherence, for a given Rabi pulsation, the reduced Allan deviation decreases for increasing free evolution time until it reaches  critical time where the width of the laser broadens the line. This behaviour results in a minimum of the reduced Allan deviation reached for this critical cycle time and depending on the laser linewidth and the Rabi pulsation. For increasing Rabi pulsation $\Omega$, this minimum Allan deviation decreases towards a limit which depends very few on the laser linewidth as it varies by less than a factor of 2 over the whole range of the considered linewidth ( $0.1 \leq \Gamma_L/2\pi \leq 100$ Hz). This is made possible by  the short interaction time with the laser, allowed by a strong Rabi pulsation $\Omega$. In the case where the experiment is not limited by the available laser power, $\Omega$ must be chosen to reach a minimum Allan deviation close to the limit but also to cause a negligeable light-shift on the atomic levels. This light-shift is evaluated in section \ref{AC} and our calculations and numerical computations show that a Rabi pulsation $\Omega=1000$ s$^{-1}$ allows to reach the Allan deviation limit to better than 1\% and to cause a negligeable light-shift.
\begin{figure}
\begin{center}
 \includegraphics[height=6.cm]{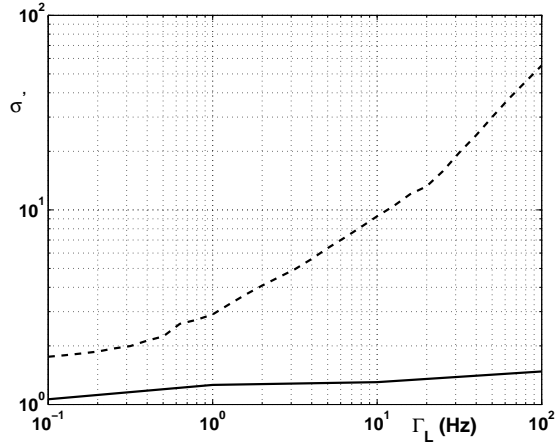}
 \caption{  Minimum reduced Allan deviation versus the laser
 linewidth $\Gamma_L/2\pi$ (Hz), for a metastable lifetime
  $\tau_m=1/\Gamma_m=1$ s. The dashed line shows the minimum achievable
  reduced deviation with a single Rabi pulse (the cycle time and Rabi pulsation 
are computed to minimize $\sigma^{'}$) . The solid line shows the minimum
  deviation achievable with a two Ramsey pulses  excitation for a fixed Rabi pulsation of $1000  \Gamma_m=$1000 s$^{-1}$ .
 \label{fig_compare_sigma}}
\end{center}
 \end{figure}
Figure \ref{fig_compare_sigma} shows this minimum reduced Allan deviation for a laser linewidth from 0.1 Hz to 100 Hz (FWHM), compared with the one reached with a Rabi  excitation scheme, where the best $\Omega$ and $T_c$ are found to reach the minimum Allan deviation, like explained in figure \ref{fig_rabi_broad}. The minimum Allan deviation expected for a  narrow laser (0.1 Hz) for a Ramsey excitation scheme  is 1.06 and  for a Rabi scheme is 1.76. These values are very close to each-other but their evolution with the width of the laser is very different in the two cases. As shown in figure \ref{fig_compare_sigma}, for a Rabi excitation scheme, the minimum Allan deviation increases with the laser linewidth to reach $\sigma^{'}=55$ for $\Gamma_L/2\pi=100$ Hz whereas it reaches 1.48 for a Ramsey scheme with the same laser linewidth. This evolution illustrates the great advantage of Ramsey two separated pulses excitation over a Rabi single pulse. 

\begin{figure}
\begin{center}
 \includegraphics[height=6.cm]{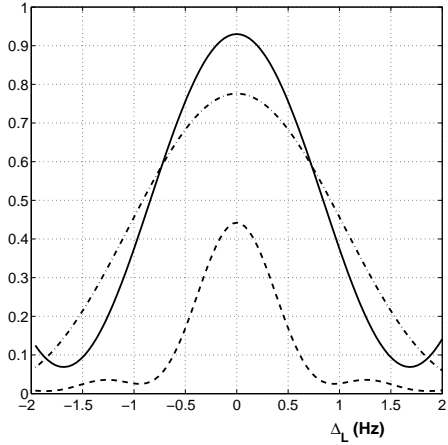}
 \caption{ Excitation profile leading to minimum Allan deviation for a  laser
 linewidth $\Gamma_L=0.1$ (Hz), for a metastable lifetime
  $\tau_m=1$ s. Solid line : profile reached for two Ramsey pulses, with $\Omega=1000 \Gamma_m$ and $T_c=313$ ms. Dashed and dot-dashed line : profiles reached for a single Rabi pulse, two configurations lead to the same $\sigma^{'}=1.76$: $\Omega=2 \Gamma_m$ and $T_c=1.093$ s (dashed line) and  $\Omega=7 \Gamma_m$ and $T_c=392$ ms (dot-dashed line).
 \label{fig_compare_profil}}
\end{center}
 \end{figure}

To give an insight of how such performances are reached,  the excitation probability profiles calculated for the conditions giving the minimum Allan deviation for a 0.1 Hz  laser linewidth are shown in figure \ref{fig_compare_profil}.
For a Ramsey scheme and the chosen Rabi pulsation of 1000 s$^{-1}$, the optimum cycle time is 313 ms and results in a 1.76 Hz wide profile ($\Delta f$) with $p_m=0.465$. For a single pulse, the minimum Allan deviation can be reached by two different Rabi pulses. One with $\Omega=2$ s$^{-1}$ and lasting 1.093 s gives rise to a narrow ($\Delta f=0.86$ Hz) but few excited profile ($p_m=0.22$). Another one with $\Omega=7$ s$^{-1}$ lasting 392 ms leads to a broader profile ($\Delta f=2.28$ Hz) with higher excitation ($p_m=0.39$). These three profiles illustrate the compromise required between high excitation probability and narrow linewidth to reduce the Allan deviation. They  show that different conditions can reach this compromise. In any case, the Ramsey excitation scheme results in Allan deviation smaller than for a Rabi scheme. Furthermore, this method has the advantage of keeping low deviation even for laser linewidth broader than the transition.

\subsubsection{Influence of the metastable level lifetime \label{effet_tau}}
We have computed the minimum reduced Allan deviation for different metastable level lifetimes $\tau_m$, for a very narrow laser ($\Gamma_L/2\pi=0.1$ Hz) and excitation by two Ramsey pulses with $\Omega=1000$ s$^{-1}$ as this value allows to reach the limit of the Allan deviation. The results are summarized in  table \ref{tab_sigma}. The computed Allan deviations are very close for all the ion optical frequency standard candidates (between 1.4 and  $5.9\time 10^{-15}/\sqrt{\tau}$) confirming quantitatively the prediction that optical frequency standard will overtake the performances of existing microwave standards in the long run.
\begin{table}
\caption{Minimum Allan deviation expected for different optical frequency standard candidates, computed with a very narrow laser  ($\Gamma_L/2\pi=0.1$ Hz) and two Ramsey pulses with  $\Omega=1000$ s$^{-1}$.  } \label{tab_sigma}
\begin{center}
\begin{tabular}{|c|c|c|c|c|}
\hline
atom & $\tau_m$ (s) & $\lambda$ (nm)& $\sigma^{'}_{mini}$ & $\sigma_y(\tau)$  \\
\hline \hline
In$^{+}$ & 0.2   & 236 & 1.78 & $1.4\time 10^{-15}/\sqrt{\tau}$  \\
\hline
Ca$^{+}$ & 1.15  &  729 & 1.05 & $2.5\time 10^{-15}/\sqrt{\tau}$  \\
\hline
Hg$^{+}$ & 0.08 &  282 & 2.78 & $2.6\time 10^{-15}/\sqrt{\tau}$  \\
\hline
Sr$^{+}$ & 0.4 &  674 & 1.40 & $3.1\time 10^{-15}/\sqrt{\tau}$  \\
\hline
Yb$^{+}$ & 0.05 &  436 & 4.04 & $5.9\time 10^{-15}/\sqrt{\tau}$  \\
\hline

\end{tabular}
\end{center}
\end{table}

\subsubsection{Summary}

First, our studies confirm that a Ramsey excitation scheme is more appropriate than a Rabi one to take full advantage of very narrow atomic transition in the goal of building a frequency standard. They also show that the finite laser linewidth implies an optimum cycle time for a given Rabi pulsation, which can not be deduced intuitively from this linewidth as it  ranges from $T_c=120$ ms for the shortest metastable lifetime listed on table \ref{tab_sigma} ($\tau_m=0.05$ s) to 312 ms for the longest metastable level lifetime  $\tau_m=1.15$ s (Ca$^{+}$). The Allan deviation expected for a Ca$^{+}$ standard is  $2.5\time 10^{-15}/\sqrt{\tau}$ which ranks well among the other candidates for optical frequency standard.

\section{Frequency standard accuracy and precision\label{s-precision}}
Besides the  frequency stability, the other relevant parameters defining the quality of a frequency
standard are its accuracy and its precision. The standard frequency
may be shifted from the atomic resonance value by any interaction of the
atom with external fields. If this shift is constant, it only
reduces the standard accuracy but not its precision.  If this
shift varies in time or can not be evaluated exactly, the
precision is reduced also. As these effects contribute to the
uncertainty of the future frequency standard, all the interactions
of the ion with its surrounding must be controlled to minimise
and/or to maintain any shift of the clock frequency. We
evaluate these shifts for a calcium ion in order to choose the
best internal state and prepare an environment for which these shifts are minimum.

The ground state $\left| g\right>$ of the calcium ion  is
$\left|S_{1/2}\right>$ and the metastable state $\left| m\right>$ is $\left|D_{5/2}\right>$
with a measured lifetime of $1152 \pm 23$ ms \cite{knoop03} which
leads to a natural width for the clock transition of $138 \pm 3$
 mHz. We require that during the excitation of the transition by
the clock laser, all other lasers are shut off. This assures
that there are no light-shifts of the levels $S_{1/2}$ and
$D_{5/2}$ caused by the cooling lasers. The other major effects
that can shift the standard frequency are due to the local
magnetic and electric fields and to the intensity of the clock laser
itself. In the Lamb-Dicke regime, the  Doppler
effect shifts the line only by its second order contribution. We first focus on the Zeeman
effect as it governs the choice of the isotope and atomic
sublevels used for the standard.

\subsection{Zeeman effect\label{s-zeeman}}
To avoid any uncontrolled or time-varying shifts, the frequency
standard must be made as independent as possible of environmental
conditions. The first order Zeeman effect can be 
eliminated by the use of atomic Zeeman sublevels with no
projection of the total moment on the magnetic field. This can be realised by the use of 
 an isotope with a half integer nuclear spin.
The most abundant one (0.135\% in a natural sample) is $^{43}$Ca$^{+}$ with a  nuclear
spin  7/2.
\begin{figure}
\begin{center}
 \includegraphics[height=9.cm,angle=-90]{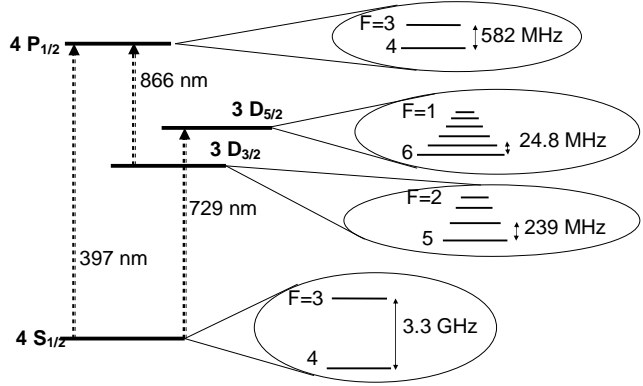}
 \caption{Hyperfine structure of the levels involved in the preparation,
 excitation and detection of $^{43}$Ca$^{+}$ ($I=7/2$) \cite{arbes94,nortershauser98}.
  \label{fig_SHF}}
\end{center}
 \end{figure}
The hyperfine structure of this isotope can be found on figure
\ref{fig_SHF}. Alternatively, the first order Zeeman effect could be eliminated on the  $^{40}$Ca$^{+}$ transition by cancellation between  $\left|S_{1/2}, m=1/2\right>\rightarrow \left|D_{5/2}, m=1/2\right> $ and  $\left|S_{1/2}, m=-1/2\right>\rightarrow \left|D_{5/2}, m=-1/2\right> $. The exact cancellation of the first order Zeeman effect to better than one hertz requires a stability of the magnetic field better than $10^{-4} \mu$T for  at least a few seconds, which seems difficult to realise. As a consequence the use of the odd isotope appears to be the easiest solution to eliminate the first order Zeeman shift from the standard frequency.

\begin{figure}
\begin{center}
 \includegraphics[height=7.cm]{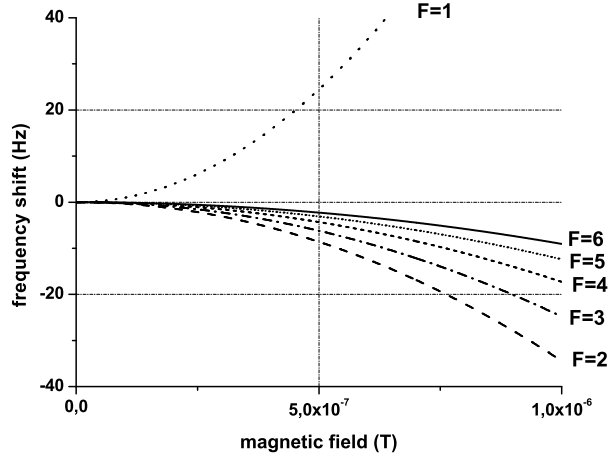}
 \caption{Zeeman shift of the
metastable hyperfine sublevels of $D_{5/2}$ versus the magnetic
field, for Zeeman sublevel $m_F=0$.
  \label{fig_zeeman}}
\end{center}
 \end{figure}

The second order Zeeman shift depends  on the
choice of the hyperfine sublevels. We calculate these shifts by
searching the eigenvalues of the Zeeman Hamiltonian for the
$\left|F, m_F \right>$ states. The Zeeman shift of the
fundamental hyperfine levels $\left|S_{1/2}, F=3{\rm \ or \ }F=4,
m_F=0\right>$ are at least 2 orders of magnitude smaller than the
shift of the metastable hyperfine levels $\left|D_{5/2},
F=1,...,6, m_F=0\right>$ and are not relevant for the choice of
the  level. Figure
 \ref{fig_zeeman} shows the quadratic Zeeman shifts for the
 different hyperfine levels of $D_{5/2}$ for sublevels $m_F=0$.
These curves illustrate the great variation of these shifts with
the hyperfine sublevel. Depending on the level involved, the
second order Zeeman effect can be as large as
 98.04 Hz/$\mu$T$^2$ for $F=1$ or reduced to -9.05 Hz/$\mu$T$^2$
 for $F=6$ (see figure \ref{fig_zeeman}). This last level should be used as a  basis to reduce the second order Zeeman effect of the
 standard. As the $\left|S_{1/2} \right>\rightarrow
\left| D_{5/2}\right>$ transition is electric-quadrupole, the selection rules
 $\Delta F=0, \pm 1, \pm 2$ imply that the fundamental sublevel
 involved in the standard should be $\left|S_{1/2}, F=4, m_F=0\right>$.

To eliminate the uncertainty due to the Zeeman effect, the local
magnetic field must be kept on the 0.1 $\mu$T level. A
 controlled magnetic field is still needed to split all the
 Zeeman sublevels and to be able to select the $\left|m_F=0\right> \rightarrow
 \left|m_F=0\right>$ transition. The two closest transitions $\left|S_{1/2}, 4, \pm1\right> \rightarrow
 \left|D_{5/2}, 6, \pm1 \right> $
  are split apart by $\pm 3.5$ kHz/$\mu$T. So a magnetic field of 0.1 $\mu$T
  allows to isolate the $m_F=0 \rightarrow
 m_F=0$ transition and can be measured by the observation of these neighbouring transitions.
  Nevertheless, such a magnetic field may not be sufficient to maintain
 a high level of
 scattered light by  the atomic system, as observed for other
 ions under study \cite{gill03}. But it is possible to recover a high level signal by
 spinning of the laser polarisation \cite{berkeland02,gill03} and
 we do not consider, at this level, this reduction of the signal as a limitation.
 Without any magnetic field applied, the local field caused by the earth and the experimental setup is of the order of $10^{-4}$ T
  and such fields can be produced by 1 A in Helmholtz coils.
 Furthermore, magnetic field fluctuations of 0.2 $\mu$T over one day have been
 observed in an unshielded environnement \cite{bize03}. As a consequence, in a
 thermalized and shielded environnement, it is  technically possible with standard current
 supplies of 1 A stabilized to the mA level, to
 compensate for the already existing magnetic field and to add the
 desired magnetic field of $0.1 \pm 0.05 \mu$T. In these conditions,
 the frequency uncertainty due to the Zeeman effect is
\begin{equation}\label{zeeman}
  B=0.1 \pm 0.05 \ \mu{\rm T} \rightarrow \delta f_Z=
  -0.09 \pm 0.09 \ {\rm Hz}
\end{equation}

 Indeed, it is possible to prepare  the atomic system in the $\left|S_{1/2}, F=4,
 m_F=0\right>$ state thanks to the property of dipole transition that
 forbids $\left|F, m_F=0 \right>\rightarrow \left|F, m_F=0\right>$ transitions. After the cooling
 stage, two lasers  polarised parallel to the magnetic field ($\Delta m_F=0$) and
 resonant with the $\left|S_{1/2}, F=4{\rm \ and\ }F=3 \right>\rightarrow \left|P_{1/2}, F=4\right>$
 transitions optically pump the system in the $\left|S_{1/2}, F=4,
 m_F=0\right>$ state in few microseconds, then ready for the probe stage.
 Actually, the cooling and optical pumping stage are not so simple
 due to a possible decay from $P_{1/2}$ to $D_{3/2}$ level (see figure \ref{fig_SHF}),
 whose lifetime is the same order of magnitude as $D_{5/2}$
 and so requires  three repumping lasers, to empty the possibly occupied
 $F=3,4,5$ levels. If the repumping
 lasers are $\sigma(=\sigma^{+}+\sigma^{-})$ polarised, the cooling and optical pumping
 remains efficient, as long as  the three repumping lasers' detunings are different from
  the two cooling ones, to avoid
 dark resonances  \cite{janik85} and as long as their polarization is spun to prevent
  pumping into dark states \cite{berkeland02}.  With its very low natural abundance, the use af such an isotope is technically challenging, but it has been shown that photoionisation processes allow to create $^{43}$Ca$^{+}$ ions even from a non-enriched calcium sample \cite{lucas04}.

\subsection{Interaction with DC electric fields and their gradients}
The second order Stark effect shifts the standard frequency
through the coupling of the levels $S_{1/2}$ and $D_{5/2}$ to all
the other atomic levels by  electric dipole interaction with any
DC or slowly varying electric fields. These fields also shift the
$D_{5/2}$ level by the coupling of its electric quadrupole moment
 to any electric field gradient. In a usual miniature spherical trap, the
confining electric field has no static component, oscillates at a
frequency  of the order of 10 MHz and its shape in the center can
be very well approximated by a quadrupole. In the exact center of
the trap there should be no oscillating field but an oscillating
field gradient. However, in a real Paul trap, patch potentials
deform the harmonic potential well created by the RF field. They
separate the minimum potentiel point from the zero RF-field point
and static bias voltages have to be applied in the three
directions to make these two points meet again and reduce any
static field to less than V/cm. This step is required to be able
to cool an ion to the Doppler limit and to reach the Lamb-Dicke
regime \cite{berkeland98} and can lead to an increase of the
static electric field gradient. Such gradient can realistically
reach 1V/mm on 1 mm (the typical diameter of a Paul-Straubel
trap). The local electric field is then the sum of the quadrupole
oscillating field that traps the ion, the bias static field lower
than V/cm and the isotropic field radiated by the vessel
considered as a blackbody. Since the frequencies of this radiated
field are far below the optical resonance of Ca$^{+}$, the field
can be taken into account by its mean-square value averaged over
all the blackbody spectrum, whose value is given by \cite{itano82}
\begin{equation}\label{BB}
 \left<E_{BB}^{2}\right>=831.9^2 \left(\frac{T}{300}\right)^4
\end{equation}
in (V/m)$^{2}$ with $T$  in Kelvin. At room temperature, this
field overtakes the static bias field resulting from compensation
of patch potentials. Nevertheless, it can be drastically reduced
by cooling the vessel, a thermalization at 77 K sets this
field below the level of V/cm,  comparable to the bias static
one.

 Thanks to a symmetry property of the second
order Stark Hamiltonian (which behaves like a second order
tensor), the Stark shift of $S_{1/2}$ is independent of the
hyperfine level and Zeeman sublevel. As a consequence, it is also
independent of the polarisation of the electric field and behaves
like a scalar (this property is true for any level with $J<1$).
An electric field couples the ground state $4 S_{1/2}$ to all the
$n P_{1/2}$ and $n P_{3/2}$ levels but in the fact, the sum of
the oscillator strength  on $4 P_{1/2}$ and $4 P_{3/2}$ is
already equal to 1 \cite{base_harvard} and there is no point
taking into account
 other couplings to $n>4$ levels. The second order Stark shift on  $4 S_{1/2}$ is
 then easily evaluated to $-9.5$ mHz/(V/cm)$^{2}$.

The Stark effect on the $D_{5/2}$ level can be split into a scalar
 term, independent on $F$ and $m_F$ and a tensorial part,
 depending on these two quantum numbers and on the angle $\theta$ between the electric
  field and the magnetic field defining the quantification axis.
The sum of all the oscillator strengths of the transitions between
$3 D_{5/2}$ and $n P_{3/2}(n\geq4)$, $n F_{5/2}(n\geq4)$ and $n
 F_{7/2}(n\geq4)$ is only 0.48 (according to the Harvard database
 \cite{base_harvard}), suggesting that there are other couplings with levels belonging to the continuum.
 Then our evaluation can only be a rough
 estimation, but it gives a correct order of magnitude. We find  $-3.9$ mHz/(V/cm)$^{2}$ for the scalar
 part and
  $+2.1$ mHz/(V/cm)$^{2}\times(3\cos^{2}\theta -1)/2$ for the tensorial
  part and associate an uncertainty as high as the value itself to
 take into account that there are missing couplings. The total frequency shift due to DC-Stark effect is then
\begin{eqnarray}\label{stark}
 \delta f_S (S_{1/2}& \rightarrow &  D_{5/2} ,  F=6,
 m_F=0)  \nonumber \\
 & = & 5.6 (\pm 4) +2.1 (\pm 2) \times\left(\frac{3\cos^2\theta-1}{2}\right) \ \ {\rm mHz/(V/cm)}^{2}
\end{eqnarray}
At room temperature, the DC Stark shift is mainly  due to the
 isotropic radiated field and is therefore
\begin{equation}\label{stark_num}
 \delta f_S (S_{1/2}\rightarrow D_{5/2}, F=6,
 m_F=0)=0.39 (\pm 0.27) \ \ {\rm Hz}
\end{equation}
If the vessel is cooled to 77 K, the contribution of the radiated
field is of the same order as the bias static field so its
direction is unknown and its amplitude of the order of 1 V/cm.
Such a field induces an uncertainty on the frequency of
\begin{equation}\label{stark_77}
 \delta f_S (77K)(S_{1/2}\rightarrow D_{5/2}, F=6,
 m_F=0)\leq 12\ \ {\rm mHz}
\end{equation}
As for the coupling of the electric quadrupole moment of the $3
D_{5/2}$ state to any electric field gradient, it depends on the
hyperfine level, its Zeeman sublevel and on the  shape as well
as on the symmetry axis of the electric potential \cite{itano00}.
The coupling strength is due to a non spherical repartition of
the electronic charge density and depends on the atomic orbitals
of the considered level. The quadrupole moment $\Theta(3D_{5/2})$
of the fine structure state can be defined as \cite{itano00}:
\begin{equation}\label{theta_def}
  \Theta(3D_{5/2})=-\frac{e}{2}\left<3D_{5/2}, \  m_J=\frac{5}{2} |3z^2-r^2|3D_{5/2}, \  m_J=\frac{5}{2}
  \right>
\end{equation}
This is calculated  by considering the electronic orbital of
$3D_{5/2}$  as pure $3d$ without any mixing with other electronic
orbitals. For a single electron atom \cite{sobelman}
\begin{equation}
 \Theta=\frac{e}{2}\left<r^2\right>\frac{2J-1}{2J+2}
\end{equation}
In our case:
\begin{equation}\label{theta}
  \Theta(3D_{5/2})=\frac{2e}{7}\left<r^2\right>_{3d}
\end{equation}
In \cite{itano00},
 the Cowan code is used to compute $\left<r^2\right>_{5d}$
for Hg$^{+}$. A good enough and simple estimation of
$\left<r^2\right>$ in alkali like ion is provided by the quantum
defect method \cite{bransden} which gives a simple relation between the energy $E_{nl}$ of the electronic level and an effective quantum number $n^{*}(l)$ by 
\begin{equation}
E_{nl}=-\frac{\widetilde{Z}^2}{n^{*2}(l)}a.u. , \widetilde{Z}=Z-N+1
\end{equation}
 $\left<r^2\right>_{3d}$ can then be calculated using the one-electron orbital properties, with $ \widetilde{Z} $ and $n^*$ instead of $Z$ and $n$. This method gives for Ca$^{+}$
\begin{equation}\label{r2}
  \left<r^2\right>_{3d}=6.6 a_0^2
\end{equation}
where $a_0$ is the Bohr radius. The energy shift of the hyperfine
sublevel $\left|F=6, m_F=0\right>$ of $3D_{5/2}$ is
\begin{equation}\label{QshiftE}
  \delta
  E=\frac{7}{11}\left(\frac{2e}{7}\left<r^2\right>_{3d}\right)\left(\frac{1}{2}\frac{\partial^2 V}
  {\partial x^2}\right)\Pi
\end{equation}
where $\Pi$ is a geometrical factor equal to $(3\cos^2 \beta -1)$
if the field has a quadrupole symmetry ($V\propto x^2+y^2-2z^2$),
$\beta$ being the angle between its symmetry axis and the
magnetic field defining the quantization axis \cite{itano00}.  The frequency
shift of the standard transition under investigation is then
\begin{equation}\label{Qshiftf}
  \delta f=8.1\times 10^{-7}\left(\frac{1}{2}\frac{\partial^2 V}
  {\partial x^2}\right)\Pi \ \ \ {\rm Hz}
\end{equation}
The hyperfine level has little influence on this shift as, for
exemple, for the level $F=2$, 7/11  is replaced by 17/35. With the
expected gradient of 1 V/mm over 1 mm, the uncertainty induced by
this effect reaches the hertz level, which is high compared to
the width of the clock transition in Ca$^{+}$. Any modification of
the patch potential,  due for exemple to the ion creation
process, alters this shift and reduces the reproductibility of
the standard. Still, this effect can be eliminated by averaging
the transition frequency measured with the magnetic field along
three perpendicular directions, as the geometrical $\Pi$ factor is
then averaged to zero \cite{itano00}. The remaining uncertainty
will then depend on the precision of the angle setting between the
three measurements. This precision  depends a lot on the vessel
design and experimental setup, and it seems difficult to
estimate this uncertainty as long as  we have not  performed
the experiment. Nevertheless, other authors \cite{gill03b} have
projected to reduce by 50 the uncertainty induced by this shift
and we assume that a  reduction by a factor of 10 is readily achievable,
which sets the uncertainty induced by the quadrupole effect to
$\pm 0.1$ Hz. At this point, it is important to mention that in
spherical traps, the field gradient is inferior to the one in 
linear traps, due to the confining geometry. As a consequence, in order to minimize the
shift induced by the gradient a
spherical trap is  preferred to a linear trap.

\subsection{Interaction with AC electric fields\label{AC}}
During the excitation of the clock transition, only one laser is applied. It can
still cause an AC-Stark shift (or light-shift) of $S_{1/2}$ and
$D_{5/2}$ by coupling them to $P_{1/2}$ and  $P_{3/2}$ by dipole
interaction or by coupling them to other Zeeman sublevels of
$D_{5/2}$ and $S_{1/2}$ by quadrupole interaction (the coupling
with $D_{3/2}$ is far less strong). The first two couplings
produce a shift proportional to the laser intensity $I_{729}$
equal to $1.1\times 10^{-4}\times I_{729}$ Hz. The laser intensity
required to produce the highest Rabi pulsation of 1000 $s^{-1}$
considered in part \ref{s-ramsey} on the $\left| S_{1/2},\ 4, \
0\right> \rightarrow \left|D_{5/2},\ 6, \ 0\right>$ transition is
0.75 $\mu$W/mm$^{2}$, which leads to a
  light-shift  caused by dipole coupling  equal to 0.08
  mHz, which is however negligeable compared to the natural width of the transition.

Light-shifts of a few kHz due to quadrupole interaction with other
Zeeman sublevels  have been measured on $^{40}$Ca$^{+}$ isotope
\cite{haffner03}.  In these experiments Rabi pulsations of 1 MHz
were used with laser detunings of the order of 1 MHz. Here we
calculate this shift in the context of the clock transition
excitation for a Rabi pulsation equal to $1000$ s$^{-1}$  and a
magnetic field of 0.1 $\mu$T. The frequency detuning required to
probe the clock transition depends on the laser linewidth and on the
 Rabi pulsation used, and is of the order of 1 Hz. By chosing $\pm
10$ Hz for this detuning the light-shift is then not
underestimated. We find an effect equal to
$\pm 6$ mHz decreasing to $\pm 0.06$ mHz if the magnetic
field is 1 $\mu$T. The sign depends on the sign of the detuning.
Two reasons make this effect very small: the small Rabi pulsation
considered for such experiments and the small detuning required
to probe the two sides of the transition (of the order of a few
Hz). With such small detuning, the couplings of $\left|S_{1/2}, 4
,0\right>$ with $\left|D_{5/2}, 6 ,2\right>$ and with
$\left|D_{5/2} ,6,-2\right>$ compensate each other (and vice-versa
for $\left|D_{5/2} ,6,0\right>$ with $\left|S_{1/2},4,\pm
2\right>$). Nevertheless, with the laser power and magnetic field
values planned for the optical clock realisation, this effect
overtakes the ones induced by dipole couplings.

\subsection{Second order Doppler shift\label{s-doppler2}}
The second order Doppler effect  shifts the frequency transition
by
\begin{equation}\label{doppler}
  \frac{\delta f_D}{f_0}=-\frac{\left<v^2\right>}{2c^2}\end{equation}
With an oscillating ion, cooled to the Doppler limit, the
velocity of the ion can be written like
$v=V_0\cos(\omega_{tr}t+\phi)$ and $\left<v^2\right>=V_0^2/2$.
This mean-square velocity is calculated by $V_0=\omega_{tr}X$ and
$X= \lambda_{tr}\sqrt{2\langle n\rangle +1}$ ({\it  cf}
\ref{s-prep}). With the  values chosen in \ref{s-prep}, the
velocity amplitude $V_0$ is equal to 0.32  m/s leading to a second order
Doppler relative shift given by
\begin{equation}
  \frac{\delta f_D}{f_0}=-2.8\times 10^{-19}
\end{equation}
In the case of the Ca$^{+}$ clock transition ($f_0=4.11 \times
10^{14}$ Hz) , the absolute shift is $0.11$ mHz, which is
negligeable in the reduction of the clock precision. This
calculation
  confirms that Doppler laser cooling is sufficient also to reduce
  the second order Doppler effect to negligeable values.

\subsection{Uncertainty budget}
Table \ref{tab_precision} gives the uncertainty budget
 expected for an atomic clock based on $^{43}$Ca$^{+}$. At room temperature,
 and with the considered magnetic field, the major
source of frequency shift and uncertainty is the Stark effect
induced by the radiated electromagnetic field. This effect is
drastically reduced in a vessel cooled to 77 K and then the major
source of uncertainty becomes
 the coupling with the field gradient through the quadrupole moment of
$D_{5/2}$ which  limits the
ultimate precision of the clock. It can be compensated by
measuring the frequency with three perpendicular directions  of
magnetic field. Nevertheless, the obtained precision will depend on the
design of the experimental setup and the ability to control the
directions of the laser propagation and magnetic field. The
projections made for all these major systematic shifts show that
an atomic frequency standard based on $\left|S_{1/2},\ 4, \ 0\right> \rightarrow
\left|D_{5/2},\ 6,\ 0\right>$ of $^{43}$Ca$^{+}$ can reach an
uncertainty of $ 4 \times 10^{-16}$, with room for improvement by better compensation of the quadrupole shift and better stabilization of the magnetic field.

\begin{table}
\caption{Uncertainty budget for the frequency transition of $\left|S_{1/2},\ 4,
\ 0\right> \rightarrow \left|D_{5/2},\ 6,\ 0\right>$ in
$^{43}$Ca$^{+}$} \label{tab_precision}
\hspace{-2.cm}\begin{tabular}{|c|c|c|c|}
\hline
effect & fields/conditions &  shift (Hz)@ 300 K & @ 77 K  \\
\hline \hline
second order Zeeman effect & 0.1 $\mu$T & $-0.09 \pm 0.09$ & $-0.09 \pm 0.09$ \\
Stark effect & radiated and bias static field   & $+0.39 \pm 0.27$  &  $\leq$ 0.012 \\
$D_{5/2}$ coupled to the field gradient & 1 V/mm$^2$ & $\pm 0.1$ & $\pm 0.1$\\

AC Stark effect @ 729 nm & 0.75 $\mu$W/mm$^2$, 0.1 $\mu$T & $\pm 0.006$& $\pm 0.006$\\
second order Doppler effect & ion cooled to the Doppler limit &
$-1 \times 10^{-4}$& $-1 \times 10^{-4}$ \\
\hline
global shift and uncertaintity  & & +0.3 $\pm 0.4$  & -0.09 $\pm$ 0.19\\
\hline
relative shift and uncertaintity & & $+7 (\pm 9)\times 10^{-16}$ & -2 ($\pm$ 4)$\times 10^{-16}$\\
\hline

\end{tabular}
\end{table}
\section{Conclusion} We have presented a theoretical evaluation of the
ultimate performances that can be expected from an optical
frequency standard based on an electric quadrupole transition of
a trapped single $^{43}$Ca$^{+}$ ion. We studied its stability
through its  Allan deviation, assuming that the signal to noise
ratio would be limited by the quantum projection noise. Our
results show that a frequency instability of $\approx 2.5\times
10^{-15}/\sqrt{\tau}$ can be expected.
 We also show that a Ramsey excitation scheme  allows to take advantage of a very narrow transition, even with a laser broader than this transition, whereas this is not possible with a single Rabi pulse. The minimum Allan deviation is also computed for the other ions which are candidates for an optical frequency standard and Calcium ranks well within this list.

In a second time, all the systematic frequency shifts have been estimated and the
environmental conditions studied in order to minimize the
frequency uncertainty.  This minimization is limited by the precision reached
in the successive orientation of 3 mutually perpendicular magnetic
fields to compensate the coupling of the $D_{5/2}$ quadrupole
with a field gradient. A technical challenge for the future
optical frequency standard  will be to point these three
perpendicular magnetic fields and to reduce the field gradient.
In this context, a miniature spherical trap is more appropriate
than a linear one to a frequency standard. Our projections  show
that with a first step  alignement and a cooled vessel, a standard based on
$\left|S_{1/2},\ 4, \ 0\right> \rightarrow \left|D_{5/2},\ 6,\
0\right>$ of $^{43}$Ca$^{+}$ can reach an uncertainty of $ 4
\times 10^{-16}$, an order of magnitude smaller than the most
precise actual microwave frequency standard \cite{bauch03}.

\subsection*{Acknowledgement}
The authors would like to  thank F. Schmidt-Kaler for very
helpful discussions. Our project has been  supported by the
Bureau National de M\'etrologie.
%\bibliography{/home/caroline/article_horlge03/etape1/bibliotrap}

\end{document}